\documentclass[11pt]{article}
\pdfoutput=1
\usepackage{amsbsy,amsmath,amsthm,amssymb,graphicx,verbatim}
\usepackage{setspace, float}
\usepackage[longnamesfirst,sort]{natbib}

\usepackage{multirow}
\usepackage{subcaption}

\usepackage{geometry}
\usepackage{color,soul}

\usepackage{booktabs}

\theoremstyle{remark}

\newfont{\msbm}{msbm10 at 11pt}

\begin{document}
\onehalfspacing

\title{Bayesian model selection on linear mixed-effects models for comparisons between multiple treatments and a control}

\author{Lei Gong \\ Department of Statistics \\ University of California, Riverside \\ {\tt lei.gong@email.ucr.edu} \and
James M. Flegal \\ Department of Statistics \\ University of California, Riverside \\ {\tt jflegal@ucr.edu} \and
Stephen R. Spindler \\ Department of Biochemistry \\ University of California, Riverside \\ {\tt spindler@ucr.edu} \and
Patricia L. Mote \\ Department of Biochemistry \\ University of California, Riverside \\ {\tt mote@ucr.edu}}

\date{\today}

\maketitle

\begin{abstract}

We propose a novel Bayesian model selection technique on linear mixed-effects models to compare multiple treatments with a control. A fully Bayesian approach is implemented to estimate the marginal inclusion probabilities that provide a direct measure of the difference between treatments and the control, along with the model-averaged posterior distributions. Default priors are proposed for model selection incorporating domain knowledge and a component-wise Gibbs sampler is developed for efficient posterior computation. We demonstrate the proposed method based on simulated data and an experimental dataset from a longitudinal study of mouse lifespan and weight trajectories.

\end{abstract}

\newpage

\section{Introduction} \label{sec:intro}

Experiments are run by researchers in biology, medicine, and various other scientific fields, to compare multiple treatments with a control or standard treatment. Often these studies are conducted over a period of time and result in unbalanced repeated measured data that is commonly analyzed by the linear mixed-effects model (LMM). The LMM allows for some subsets of the regression parameters to vary among subjects, thereby accounting for sources of natural heterogeneity in the population. It models the mean response as a combination of population characteristics (fixed-effects) that are assumed to be shared by subjects, and subject-specific characteristics (random-effects) that are unique to a particular subject. It is common to introduce a set of fixed-effects for each group to model the effect of the treatment \citep[see e.g.][]{fitz:2004}. To compare treatment groups with the control groups is, therefore, equivalent to comparing the sets of fixed-effects. Researchers are often interested in deciding which treatments are different from the control, and measuring the corresponding significance of the discrepancy. 

Standard model selection procedures can be implemented to answer these questions \citep[see e.g.][]{bolk:broo:clar:2009, fitz:2004} with certain limitations. One can select models by using hypothesis tests \citep[][]{step:busk:2005}; that is, test simpler nested models against more complex models and report corresponding p-values. Although the likelihood ratio test (LRT) is widely used to determine the contribution of a factor in a model throughout statistics, it is not recommended by \cite{pinh:bate:2006} for testing fixed-effects in LMM, because of its unreliability for small to moderate sample size. Also, when the focus is to compare multiple treatments to a control, \cite{burn:ande:2002} criticize that such a pairwise comparison as an abuse of hypothesis testing. Another extensively used approach is the information-theoretic model selection procedure that allows comparison of multiple models \citep[see e.g.][]{burn:ande:2002}. This method relies on information criteria, such as Akaike information criterion and Bayesian information criterion (BIC), that use deviance as a measure of fit with a penalization on more complex models. Instead of reporting p-values, it estimates the magnitude of difference between models in expected predictive power and uses this to make a decision as to whether a variable should be included in the model or not. The resulting dichotomous decisions overly simplify the problem, and, in our case, withhold important information on the magnitude of the difference between a treatment and the control.

Motivated by these practical challenges faced by frequentist approaches, we resort to Bayesian model selection techniques \citep[for a review see e.g.][]{geor:mccu:1997, clyd:geor:2004, kuo:mall:1998}. In the Bayesian framework, this problem can be transformed to the form of parameter estimation \citep[][]{ohar:sill:2009}. That is, estimating the marginal posterior probability that a variable should be in the model, i.e. the marginal inclusion probability, which can usually be calculated directly from the posterior inference using an Markov chain Monte Carlo (MCMC) simulation. 

There is an extensive literature on Bayesian model selection. \cite{geor:mccu:1993, gewe:1996} develop the stochastic search variable selection (SSVS) technique for linear regression models that uses a Gibbs sampler to traverse the model space. \cite{smit:kohn:1996} extend its application to nonparametric regression models and show how integrating the regression parameters is essential to reliable convergence of a Gibbs sampler. \cite{kohn:smit:chan:2001} propose a more efficient single-site Metropolis-Hastings sampler. \cite{holm:deni:mall:2002} consider selection and smoothing for a series of seemingly unrelated regressions. \cite{chen:duns:2003, kinn:duns:2007} develop variable selection for both fixed and random effects in generalized LMM. Recently, Bayesian model selection methods are extended to a series of spatially linked regression for functional magnetic resonance imaging analysis \citep[see e.g.][]{lee:jone:caff:bass:2014, smit:fahr:2007}. However, we are unaware of any work to extend Bayesian model selection on LMM to compare multiple treatments with a baseline. 

In this article, we develop a novel Bayesian model selection approach on LMM that accommodates and compares multiple treatment effects. The method includes a re-parameterization of the fixed-effects of each treatment that attributes part of the effect to a baseline for direct measure of the difference between a treatment and the control. A modification of the fractional prior \citep[][]{smit:kohn:1997} is proposed to undertake model selection and averaging, which is also related to Zellner's g-prior \citep{zell:1986}.  The proposed prior incorporates information on subjects within the same group, which is critical to developing an efficient component-wise Gibbs sampler.  This Bayesian paradigm provides practitioners with an intuitive understanding of the significance of each treatment through the marginal inclusion probability, which is unaccessible using existing techniques.  

Our work is motivated by a longitudinal experiment of mouse lifespan and weight trajectories \citep[see e.g.][]{spin:mote:2013a, spin:mote:2013b, spin:mote:2014a, spin:mote:2014b, spin:mote:2014c} that aims to study how different treatments affect lifetime weight trajectories and identify potential longevity therapeutics. This application provides both a clear demonstration of our approach, and an example of enabling researchers to obtain previously unavailable information.  However, the method itself is more general and applicable to most experiments that are interested in comparing multiple treatments to a control. 

\subsection{Experimental Data} \label{sec:data}

The experimental data is from a longitudinal study of the lifespan and the weight trajectories of an F1 hybrid mouse \citep[see e.g.][]{spin:mote:2013a, spin:mote:2013b, spin:mote:2014a, spin:mote:2014b, spin:mote:2014c}. The study is part of a compound screening program designed to identify potential longevity therapeutics, and it was approved by the Institutional Animal Care and Use Committee at the University of California, Riverside. It utilized an unbalanced statistical design to compare the lifespan and the weight trajectories of multiple treatment groups to that of one larger control group \citep[][]{ jesk:fleg:spin:2014}. The disposition of dietary calories between body weight and metabolic energy appears to be a key to lifespan determination. In this article, we use a part of the dataset that recorded mouse body weight changes during the course of the experiment.

In the study, 2266 male C3B6F1 mice were initially fed a chow diet \textit{ad libitum}. At 12 months of age (Day 365), 297 mice were shifted to daily feeding with 13.3 kcal/day/mouse of the control diet (Diet No.99), and the rest were shifted to control diet supplemented with one of 56 chemical, pharmaceutical, or nutraceuticals agents or combination of agents. All mice were fed daily and weighted bimonthly, but the number of mice progressively declined as the study progressed due to the onset of various age-related pathologies. The data are censored at extreme old age (Day 1369), when less than 1\% of the mice remained. 

The control and drug-treated mice gradually lost weight after the shift to the defined diets, which provided about 10\% less than the \textit{ad libitum} number of calories, to ensure the mice consumed all their food. Our main interest is to determine which supplemented diets significantly affected the lifetime weight trajectories. That is, researchers are interested in whether any deviation from the trajectory of the control group (Diet No.99) is statistically significant and is caused by dietary additions.

The rest of the paper is organized as follows. Section~\ref{sec:model} formally introduces the Bayesian variable selection methodology. It outlines the re-parameterization of a LMM, prior specification, MCMC sampling schemes, and stopping criterion utilized. A simulation study is also detailed to evaluate the performance of the proposed method. Section~\ref{sec:app} contains the empirical results from the analysis of the motivating example. Section~\ref{sec:disc} concludes with a discussion.

\section{Model Selection on Linear Mixed-effects Models} \label{sec:model}

In general, suppose that we have $n$ subjects from $G$ experimental groups under study, each with $n_i$ observations taken repeatedly over time, $i = 1, \cdots, n$, and let $\pmb{y}_{i} = (y_{i,1}, \cdots, y_{i,n_{i}})^T$ denote the response vector for the $i$-th subject. Assume the $i$-th subject is from the $g$-th group, for $i = 1, \cdots, n$, $g = 1, \cdots, G$, let $X_i$ and $Z_i$ be two $n_i \times p$ design matrices, then a LMM \citep[][]{fitz:2004, mccu:neld:1989} is denoted as 
\begin{equation} \label{eq:mixed}
\pmb{y}_i = X_i \pmb{\alpha}_g + Z_i \pmb{b}_i + \pmb{\epsilon}_i, \hspace{4mm} \pmb{\epsilon}_i \sim N_{n_i} (\pmb{0}, \sigma^2 I),
\end{equation}
where $\pmb{\alpha}_g = (\alpha_{g,0}, \cdots, \alpha_{g,p-1})^T$ are the fixed effects shared by subjects in the $g$-th experimental group. Further, denote $\pmb{b}_i = (b_{i,0}, \cdots, b_{i, p-1})^T \sim N_p (\pmb{0}, \lambda_D^{-1} I)$ as the random effects that are unique to the $i$-th subject, and hence we allow subject specific trajectories. 

Note that, among the $G$ groups, there is one control group and $G-1$ treatment groups. Without loss of generality, let us assume the $G$-th group is the control group, and $g = 1, \cdots, G-1$ are the treatment groups. A primary goal for many experiments is to determine which alternative treatments significantly differ from the control group. To this end, we propose a re-parameterization of the fixed effects $\pmb{\alpha}_g$'s in \eqref{eq:mixed}, $g= 1, \cdots G$. Let $W_i$, $X_i$ and $Z_i$ be three $n_i \times p$ design matrices, the re-parameterized model is denoted as, for $i = 1, \cdots, n$, $g = 1, \cdots, G$, 
\begin{equation} \label{eq:lme}
\pmb{y}_i = W_i \pmb{\alpha} + X_i \pmb{\beta}_g + Z_i \pmb{b}_i + \pmb{\epsilon}_i, \hspace{4mm} \pmb{\epsilon}_i \sim N_{n_i} (\pmb{0}, \sigma^2 I),
\end{equation}
where $\pmb{b}_i = (b_{i,0}, \cdots, b_{i, p-1})^T \sim N_p (\pmb{0}, \lambda_D^{-1} I)$ are the random effects as in \eqref{eq:mixed}, and $\pmb{\alpha} = (\alpha_{0}, \cdots, \alpha_{p-1})^T$ are the fixed effects of the control group, $\pmb{\beta}_g = (\beta_{g,0}, \cdots, \beta_{g,p-1})^T$ are the fixed effects modeling the difference between the $g$-th group and the control group. That is, the group effect $\pmb{\alpha}_g$ in \eqref{eq:mixed} is re-written as $\pmb{\alpha} + \pmb{\beta}_g$ in \eqref{eq:lme}, for $g = 1, \cdots, G$. Also, it is straightforward to see, as the baseline, $\pmb{\beta}_G = (0, \cdots, 0)^T$ for the control group.

Under the re-parameterization, the detection of significant treatments is equivalent to the identification of nonzero $\pmb{\beta}_g$'s. To this end, we introduce 0/1 binary indicators $\pmb{\gamma}_g = (\gamma_{g, 0}, \cdots, \gamma_{g, p-1})^T$, $g = 1, \cdots, G$, such that $\beta_{g, j} = 0$ if $\gamma_{g, j} = 0$ and $\beta_{g, j} \neq 0$ if $\gamma_{g, j} = 1$. The $\gamma_{g, j}$ is used to indicate whether the fixed effect on the $j$-th predictor of the $g$-th group differs from that fixed effect of the control group. Given $\pmb{\gamma}_g$, let $\pmb{\beta}_g (\pmb{\gamma}_g)$ be the vector of nonzero fixed effects and $X_i (\pmb{\gamma}_g)$ be the corresponding design matrix. Then, the model \eqref{eq:lme} can be written as, for $i = 1, \cdots, n$, $g = 1, \cdots, G$,
\begin{equation} \label{eq:ssvs}
\pmb{y}_i = W_i \pmb{\alpha} + X_i (\pmb{\gamma}_g) \pmb{\beta}_g (\pmb{\gamma}_g) + Z_i \pmb{b}_i + \pmb{\epsilon}_i.
\end{equation}

This formulation allows us to look at the problem from the Bayesian SSVS perspective \citep{geor:mccu:1993}. The SSVS searches for models having high posterior probability by traversing the model space using MCMC techniques, and, thus, identifies subsets of predictors with nonzero coefficients. Moreover, it allows us to calculate the posterior distributions of the parameters by marginalizing over the other variables. In this way, the marginal inclusion probability can be obtained as a direct measure of the significance of each treatment.

Note that, \eqref{eq:ssvs} is a very general setting that is applicable to a wide range of applications. It is possible to impose specific structures on $\gamma$ to suit different scenarios to further simplify the modeling procedure. For example, given the setups of the motivating experiment, it is reasonable to assume a common intercept for all groups, since the mice were on the same diet at the first measurement; that is, $\beta_{g, 0} = 0$, for $g = 1, \cdots, G$. Therefore, given that the primary goal is to compare treatment groups to the baseline group $G$, it is desirable to impose the following settings on $\gamma$,
\[
\gamma_{1, 0} = \cdots = \gamma_{G-1, 0} = \gamma_{G, 0} = 0,
\]
\[
\pmb{\gamma}_G = (\gamma_{G, 0}, \cdots, \gamma_{G, p-1})^T = (0, \cdots, 0)^T.
\]

\subsection{Prior Specification} \label{sec:prior}

A proper prior must be placed on the nonzero coefficients $\pmb{\beta}_g (\pmb{\gamma}_g)$ to undertake model averaging \citep[see e.g.][]{mitc:beau:1988, smit:kohn:1996, geor:mccu:1993, kohn:smit:chan:2001}. In particular, \cite{kohn:smit:chan:2001, smit:fahr:2007} propose a conditional prior for the coefficients by setting it proportional to a fraction of the likelihood. This fractional prior is related to the g-prior in \cite{zell:1986}, and is located and scaled in line with the information from the likelihood. We propose a modification of this idea to accommodate multiple subjects within a group by setting $\pi \left(\pmb{\beta}_g (\pmb{\gamma}_g) | y, \pmb{\alpha}, \pmb{\gamma}_g, b, \sigma^2 \right) \propto \Pi_{i \in g} \ p \left( \pmb{y}_i | \pmb{\alpha}, \pmb{\beta}_g (\pmb{\gamma}_g), \pmb{\gamma}_g, \pmb{b}_i, \sigma^2 \right)^{1/n_i}$, so that
\begin{equation} \label{eq:frac}
\pmb{\beta}_g (\pmb{\gamma}_g) | y, \pmb{\alpha}, \pmb{\gamma}_g, b, \sigma^2 \sim N \left( \hat{\pmb{\beta}}_g (\pmb{\gamma}_g), \sigma^2 \left( \sum_{i \in g} {1\over n_i} X_i^T(\pmb{\gamma}_g) X_i(\pmb{\gamma}_g) \right)^{-1} \right),
\end{equation}
where $\hat{\pmb{\beta}}_g (\pmb{\gamma}_g) = \left(\sum_{i \in g} {1\over n_i} X_i^T(\pmb{\gamma}_g) X_i(\pmb{\gamma}_g) \right)^{-1}  \left( \sum_{i \in g} {1\over n_i} X_i^T(\pmb{\gamma}_g) \left( \pmb{y}_i - W_i \pmb{\alpha} - Z_i \pmb{b}_i \right) \right)$, and $\sum_{i \in g}$ stands for summation over all the subjects that belong to the $g$-th group. 

This prior is proportional to the variance of the least squares estimate of $\beta$, and enjoys a number of attractive properties as pointed out by \cite{kohn:smit:chan:2001}. The prior \eqref{eq:frac} is rescaled automatically if the design matrix $X$ or the data $y$ is rescaled because of its structure and the presence of $\sigma^2$. Moreover, this prior is invariant to location changes in X and y given the basis term $(1, \cdots, 1)^T$ is included in $X$. Also it is data-based since $\hat{\pmb{\beta}}_g (\pmb{\gamma}_g)$ depends on $y$, which allows proper centering of $\beta$. 

We consider the prior on $\gamma$ to be $\pi (\pmb{\gamma}_g | \pi_g) = \prod_{j = 0}^{p-1} \pi (\gamma_{g, j} | \pi_g), g = 1, \cdots, G$, where $\pi (\gamma_{g, j} | \pi_g) \sim Bernoulli (\pi_g)$ and $\pmb{\pi} = (\pi_1, \cdots, \pi_G)^T$ is a vector of hyper-parameters that represents prior knowledge for every groups. Intuitively, $\pi_g$ is the probability that researchers believe the $g$-th group is significantly different from the control group before conducting the experiment. For instance, we find a sensible setting, when there is little prior knowledge of the effects of the treatments, to be letting $\pi_G = 0$ for the control group, and $\pi_1 = \cdots = \pi_{G-1} = 0.5$ for the $G - 1$ treatment groups.  We assume standard priors in Bayesian hierarchical models \citep[see e.g.][]{smit:kohn:1996, gelm:carl:ster:rubi:2004, john:jone:2010} for the rest of the parameters, i.e. $\pmb{\alpha}, b, \lambda_D, \sigma^2$,
\[
\pmb{\alpha} | \pmb{d}_3, d_4 \sim N_p (\pmb{d}_3, d_4^{-1})
\]
\[
\pmb{b}_i | \lambda_D \sim N_p (\pmb{0}, \lambda_D^{-1} I), i = 1, \cdots, n
\]
\[
\lambda_D | d_1, d_2 \sim \Gamma (d_1, d_2)
\]
\[
\pi (\sigma^2) \propto 1 / \sigma^2
\]
where $d_1, d_2, \pmb{d}_3, d_4$ are hyper-parameters to be specified.

\subsection{Posterior Inference} \label{sec:posterior}

Combining the priors and likelihoods, the full joint posterior density for $\theta = (\pmb{\alpha}, \beta, \gamma, b, \sigma^2, \lambda_D)$ is characterized by
\begin{equation} \label{eq:post}
\begin{split}
q( \pmb{\alpha}, \beta, \gamma, b, \sigma^2, \lambda_D | y) \propto & \left[ \prod_{g = 1}^G \left[ \prod_{i \in g} p\left( \pmb{y}_i | \pmb{\alpha}, \pmb{\beta}_g, \pmb{\gamma}_g, \pmb{b}_i, \sigma^2 \right) \pi \left( \pmb{b}_i | \lambda_D \right) \right] \pi \left( \pmb{\beta}_g | \pmb{\alpha}, \pmb{\gamma}_g, b, \sigma^2 \right) \pi \left( \pmb{\gamma}_g \right) \right] \\
& \times \pi (\pmb{\alpha}) \pi (\lambda_D) \pi (\sigma^2).
\end{split}
\end{equation}

This distribution has a complex form which we cannot sample from directly; instead, we resort to MCMC methodology for the posterior inference and employ a component-wise strategy \citep[][]{john:jone:neat:2013}. Specifically, we propose a component-wise Gibbs sampler for posterior computation. To this end, we need the full conditional posterior distributions of each of the parameters in $\theta$ to update the Markov chain. The derivation of the full conditional posterior distributions follows from \eqref{eq:post} using straightforward algebraic route.% (see Appendix~\ref{app:post}).

Schematically, we can set up a six-variable component-wise Gibbs sampler; that is, if we let $\theta = (\gamma, \beta, \pmb{\alpha}, \sigma^2, b, \lambda_D)$ be the current state and $\theta' = (\gamma', \beta', \pmb{\alpha}', (\sigma^2)', b', \lambda_D')$ be the future state, we iteratively sample from the full conditional posterior distributions to update the chain,
\begin{equation*}
\begin{split}
(\gamma, \beta, \pmb{\alpha}, \sigma^2, b, \lambda_D) & \rightarrow  (\gamma', \beta, \pmb{\alpha}, \sigma^2, b, \lambda_D) \rightarrow   (\gamma', \beta', \pmb{\alpha}, \sigma^2, b, \lambda_D) \rightarrow  (\gamma', \beta', \pmb{\alpha}', \sigma^2, b, \lambda_D) \\
& \rightarrow  (\gamma', \beta', \pmb{\alpha}', (\sigma^2)', b, \lambda_D) \rightarrow  (\gamma', \beta', \pmb{\alpha}', (\sigma^2)', b', \lambda_D) \rightarrow  (\gamma', \beta', \pmb{\alpha}', (\sigma^2)', b', \lambda_D').\\
\end{split}
\end{equation*}

\begin{itemize}
\item[\bf{Step 1.}] The transition $\gamma \rightarrow \gamma{'}$ consists of $G \times p$ steps,
\begin{equation*}
\begin{split}
(\gamma_{1, 0}, \gamma_{1, 1}, \cdots, \gamma_{1, p-1}, \cdots, \gamma_{G, 0}, \cdots, \gamma_{G, p-1}) & \rightarrow  (\gamma{'}_{1, 0}, \gamma_{1, 1}, \cdots, \gamma_{1, p-1}, \cdots, \gamma_{G, 0}, \cdots, \gamma_{G, p-1}) \\
& \rightarrow  (\gamma{'}_{1, 0}, \gamma{'}_{1, 1}, \cdots, \gamma_{1, p-1}, \cdots, \gamma_{G, 0}, \cdots, \gamma_{G, p-1}) \\
& \vdots \\
& \rightarrow  (\gamma{'}_{1, 0}, \gamma{'}_{1, 1}, \cdots, \gamma{'}_{1, p-1}, \cdots, \gamma{'}_{G, 0}, \cdots, \gamma{'}_{G, p-1}).
\end{split}
\end{equation*}

From the Appendix%~\ref{app:post}
, we have, for $g = 1, \cdots, G$ and $j = 0, \cdots, p-1$,  
\begin{equation}\label{eq:gam}
\begin{split}
& q ( \gamma_{g, j} | \pmb{\alpha}, \pmb{\gamma}_{- (g, j)}, b, \sigma^2, y ) \propto  \pi_{g}^{\gamma_{g, j}} ( 1- \pi_g)^{1- \gamma_{g, j}} \left( {|\sum_{i \in g} {1\over n_i} X_i^T (\pmb{\gamma}_g) X_i (\pmb{\gamma}_g) | \over |\sum_{i \in g} (1 + {1\over n_i}) X_i^T (\pmb{\gamma}_g) X_i (\pmb{\gamma}_g)|} \right)^{1\over2} \\
& \times \exp \Bigg\{-{1 \over 2\sigma^2} \Bigg[ \sum_{i \in g} \pmb{\phi}_i^T \pmb{\phi} + \bigg(\sum_{i \in g} {1\over n_i} X_i^T (\pmb{\gamma}_g) \pmb{\phi}_i\bigg)^T \bigg(\sum_{i \in g} {1\over n_i} X_i^T (\pmb{\gamma}_g) X_i (\pmb{\gamma}_g)\bigg)^{-1} \bigg(\sum_{i \in g} {1\over n_i} X_i^T (\pmb{\gamma}_g) \pmb{\phi}_i\bigg) \\
& \hspace{4em} - \bigg(\sum_{i \in g} (1+{1\over n_i}) X_i^T (\pmb{\gamma}_g) \pmb{\phi}_i\bigg)^T \bigg(\sum_{i \in g} (1 + {1\over n_i}) X_i^T (\pmb{\gamma}_g) X_i (\pmb{\gamma}_g)\bigg)^{-1} \bigg(\sum_{i \in g} (1+{1\over n_i}) X_i^T (\pmb{\gamma}_g) \pmb{\phi}_i\bigg)\Bigg] \Bigg\},
\end{split}
\end{equation}
where $\pmb{\gamma}_{- (g, j)} = (\gamma_{g, 0}, \cdots, \gamma_{g, j-1}, \gamma_{g, j+1}, \cdots, \gamma_{g, p-1})^T$ and $\pmb{\phi}_i = \pmb{y}_i - W_i \pmb{\alpha} - Z_i \pmb{b}_i$. 

At each step, an update is simulated from $\gamma_{g, j}{'} \sim q ( \gamma_{g, j} | \pmb{\alpha}, \pmb{\gamma}_{- (g, j)}, b, \sigma^2, y )$. Since $\gamma_{g, j}$ is binary, i.e. $\gamma_{g, j} \in \{ 0, 1\}$, the conditional posterior distribution $q ( \gamma_{g, j} | \pmb{\alpha}, \pmb{\gamma}_{- (g, j)}, b, \sigma^2, y )$ is easily normalized by evaluating \eqref{eq:gam} for $\gamma_{g, j} = 0$ and $\gamma_{g, j} = 1$.

\item[\bf{Step 2.}] The transition $\beta \rightarrow \beta^{'}$ consists of $G$ steps,
\begin{equation*}
\begin{split}
(\pmb{\beta}_{1}, \pmb{\beta}_2, \cdots, \pmb{\beta}_{G}) & \rightarrow  (\pmb{\beta}_{1}{'}, \pmb{\beta}_2, \cdots, \pmb{\beta}_{G}) \\
& \rightarrow  (\pmb{\beta}_{1}{'}, \pmb{\beta}_2{'}, \cdots, \pmb{\beta}_{G}) \\
& \vdots \\
& \rightarrow  (\pmb{\beta}_{1}{'}, \pmb{\beta}_2{'}, \cdots, \pmb{\beta}_{G}{'}).
\end{split}
\end{equation*}

At each step, an update is simulated from a $p$-dimensional multivariate normal distribution,
\begin{equation}\label{eq:beta}
\begin{split}
\pmb{\beta}_g{'}(\pmb{\gamma}_g) & \sim \ q( \pmb{\beta}_g(\pmb{\gamma}_g) | \pmb{\alpha}, \pmb{\gamma}_g, b, \sigma^2, y ) \\
%& \propto \left[ \prod_{i \in g} p(\pmb{y}_i | \pmb{\alpha}, \pmb{\beta}_g, \pmb{\gamma}_g, \pmb{b}_i, \sigma^2)  \right] \pi(\pmb{\beta}_g(\pmb{\gamma}_g) | \pmb{\alpha}, \pmb{\gamma}_g, b, \sigma^2) \\
& \sim N_{\sum_{j = 0}^{p-1} \gamma_{g, j}} \left( V_1^{-1} \left[ {1\over \sigma^2} \sum_{i \in g} X_i^T (\pmb{\gamma}_g) (\pmb{y}_i - W_i\pmb{\alpha} - Z_i \pmb{b}_i) \right], V_1^{-1} \right),
\end{split}
\end{equation}
where $V_1 = {1\over \sigma^2} \sum_{i \in g} (1 + {1\over n_i}) X_i^T (\pmb{\gamma}_g) X_i (\pmb{\gamma}_g)$.

\item[\bf{Step 3.}] Consider updating $\pmb{\alpha}$ where the update is simulated from a $p$-dimensional multivariate normal distribution,
\begin{equation}\label{eq:alpha}
\begin{split}
\pmb{\alpha}{'} & \sim \ q( \pmb{\alpha} | \beta, \gamma, b, \sigma^2, y ) \\
%& \propto \left[ \prod_{g = 1}^G \prod_{i \in g} p\left( \pmb{y}_i | \pmb{\alpha}, \pmb{\beta}_g, \pmb{\gamma}_g, \pmb{b}_i, \sigma^2 \right) \right]  \pi (\pmb{\alpha})  \\
& \sim N_p \Bigg( V_2^{-1} \Bigg[ {1\over \sigma^2} \sum_{g=1}^G \Bigg( \sum_{i \in g} W_i^T \bigg(\pmb{y}_i - X_i (\pmb{\gamma}_g) \pmb{\beta}_g (\pmb{\gamma}_g) - Z_i \pmb{b}_i\bigg) \\
& \quad + \bigg(\sum_{i \in g} {1\over n_i}X_i^T(\pmb{\gamma}_g) W_i \bigg)^T\bigg(\sum_{i \in g} {1\over n_i} X_i^T(\pmb{\gamma}_g) X_i(\pmb{\gamma}_g) \bigg)^{-1} \bigg(\sum_{i \in g} {1\over n_i}X_i^T(\pmb{\gamma}_g) (\pmb{y}_i - X_i(\pmb{\gamma}_g) \pmb{\beta}_g(\pmb{\gamma}_g) - Z_i \pmb{b}_i) \bigg)\Bigg) \\
& \hspace{4em} + d_4 \pmb{d}_3\Bigg] , V_2^{-1} \Bigg),
\end{split}
\end{equation}
where $$V_2 = {1 \over \sigma^2} \sum_{g = 1}^G \Bigg[ \sum_{i \in g} W_i^T W_i + \bigg(\sum_{i \in g} {1\over n_i}X_i^T(\pmb{\gamma}_g) W_i \bigg)^T\bigg(\sum_{i \in g} {1\over n_i} X_i^T(\pmb{\gamma}_g) X_i(\pmb{\gamma}_g) \bigg)^{-1} \bigg(\sum_{i \in g} {1\over n_i}X_i^T(\pmb{\gamma}_g) W_i \bigg) \Bigg] + d_4.$$

\item[\bf{Step 4.}] Consider updating $\sigma^2$. At each step, an update is simulated from a Inverse-Gamma distribution, %i.e. $(\sigma^2){'} \sim q( \sigma^2 | \pmb{\alpha}, \beta, \gamma, b, y )$.
\begin{equation}\label{eq:sigma}
\begin{split}
(\sigma^2){'} & \sim \ q( \sigma^2 | \pmb{\alpha}, \beta, \gamma, b, y ) \\
& \sim Inv\text{-}Gamma \Bigg( {1\over2} (N + \sum_{g=1}^G \sum_{j = 0}^{p-1} \gamma_{g, j} ), \\ 
 {1\over2} \sum_{g = 1}^G \Bigg[ & \sum_{i \in g} \bigg(\pmb{y}_i - W_i \pmb{\alpha} - X_i (\pmb{\gamma}_g) \pmb{\beta}_g (\pmb{\gamma}_g) - Z_i \pmb{b}_i\bigg)^T \bigg(\pmb{y}_i - W_i \pmb{\alpha} - X_i (\pmb{\gamma}_g) \pmb{\beta}_g (\pmb{\gamma}_g) - Z_i \pmb{b}_i\bigg) \\
& + \bigg[\pmb{\beta}_g - \bigg(\sum_{i \in g} {1\over n_i} X_i^T(\pmb{\gamma}_g) X_i(\pmb{\gamma}_g) \bigg)^{-1}\bigg(\sum_{i \in g}{1\over n_i} X_i^T (\pmb{\gamma}_g) \pmb{\phi}_i \bigg)  \bigg]^T \bigg(\sum_{i \in g} {1\over n_i} X_i^T(\pmb{\gamma}_g) X_i(\pmb{\gamma}_g)\bigg)^{-1} \\
& \hspace{1em} \bigg[\pmb{\beta}_g - \bigg(\sum_{i \in g} {1\over n_i} X_i^T(\pmb{\gamma}_g) X_i(\pmb{\gamma}_g) \bigg)^{-1}\bigg(\sum_{i \in g}{1\over n_i} X_i^T (\pmb{\gamma}_g) \pmb{\phi}_i \bigg)  \bigg] \Bigg] \Bigg),
\end{split}
\end{equation}
where $N = \sum_{g=1}^G \sum_{i \in g} n_i$. Note that, we denote $Inv\text{-}Gamma (\alpha, \beta) = {\beta^{\alpha} \over \Gamma(\alpha)} x^{-\alpha - 1} \exp \left({-\beta \over x} \right)$, for $x \in (0, \infty)$, and $\alpha, \beta > 0$.

\item[\bf{Step 5.}] The transition $b \rightarrow b{'}$ consists of $n$ steps, 
\begin{equation*}
\begin{split}
(\pmb{b}_1, \pmb{b}_2, \cdots,\pmb{b}_n) & \rightarrow  (\pmb{b}_{1}{'}, \pmb{b}_2, \cdots, \pmb{b}_n) \\
& \rightarrow  (\pmb{b}_{1}{'}, \pmb{b}_2{'}, \cdots, \pmb{b}_{n}) \\
& \vdots \\
& \rightarrow  (\pmb{b}_{1}{'}, \pmb{b}_2{'}, \cdots, \pmb{b}_{n}{'}).
\end{split}
\end{equation*}

At each step, assuming the $i$-th subject is from the $g$-th group, an update is simulated from a $p$-dimensional multivariate normal distribution,
\begin{equation}\label{eq:b}
\begin{split}
\pmb{b}_i{'} & \sim \ q( \pmb{b}_i | \pmb{\alpha}, \pmb{\beta}_g, \pmb{\gamma}_g, \sigma^2, \lambda_D, \pmb{y}_i) \\
%& \propto p\left( \pmb{y}_i | \pmb{\alpha}, \pmb{\beta}_g, \pmb{\gamma}_g, \pmb{b}_i, \sigma^2 \right) \pi \left( \pmb{b}_i | \lambda_D \right) \\
& \sim N_{p} \Bigg( V_3^{-1} {1\over \sigma^2} \Bigg[ {1\over n_i}Z_i^T X_i(\pmb{\gamma}_g) \bigg(\sum_{j \in g} {1\over n_j} X_j^T(\pmb{\gamma}_g) X_j(\pmb{\gamma}_g) \bigg)^{-1} \bigg(\sum_{\substack{{j \in g}\\{j \neq i}}} {1\over n_j} X_j^T(\pmb{\gamma}_g) \pmb{\phi}_j + {1\over n_i}X_i^T(\pmb{\gamma}_g)\pmb{\phi}_i \bigg) \\
& \hspace{6em} + Z_i^T \bigg(\pmb{y}_i - W_i \pmb{\alpha} - (1 + {1\over n_i}) X_i (\pmb{\gamma}_g) \pmb{\beta}_g (\pmb{\gamma}_g) \bigg) \Bigg], V_3^{-1} \Bigg),
\end{split}
\end{equation}
where $V_3 = {1\over \sigma^2} Z_i^T Z_i + \lambda_D I + {1\over \sigma^2} {1\over n_i} Z_i^T X_i(\pmb{\gamma}_g) \bigg(\sum_{j \in g} {1\over n_j} X_j^T(\pmb{\gamma}_g) X_j (\pmb{\gamma}_g)\bigg)^{-1} X_i^T(\pmb{\gamma}_g)Z_i$.

\item[\bf{Step 6.}] Finally, consider updating $\lambda_D$.  At each step, an update is simulated from a Gamma distribution,
\begin{equation}\label{eq:lambda}
\begin{split}
\lambda_D{'} & \sim \ q( \lambda_D | b ) \\
& \sim \Gamma ({np \over 2} + d_1, \ {1\over 2} \sum_{g=1}^G \sum_{i \in g} \pmb{b}_i^T \pmb{b}_i + d_2).
\end{split}
\end{equation}

\end{itemize}

The posterior inference on model parameters can be carried out using the MCMC samples. Models with high posterior probability can be identified as those appearing most often in the MCMC output. One posterior quantity of interest is the marginal inclusion probability, i.e. $\text{Pr}(\gamma_{g, j} = 1 | y)$, $g = 1, \cdots, G$, $j = 0, \cdots, p-1$, which can be calculated using the proportion of draws in which $\gamma_{g, j}$ is non-zero. %Treatment groups with higher inclusion probability effectively means that these treatments are significantly different from the control group. 
It provides a direct measure of the significance of $\beta_{g, j}$, which remains challenging for the current frequentist methods. It, therefore, allows researchers for straightforward understanding of the significance of each treatment. Also, if needed, one may classify a treatment effect such that $\text{Pr}(\gamma_{g, j} = 1 | y) > 0.8772$ as significant or otherwise insignificant \citep[see e.g.][]{smit:fahr:2007, raft:1996, lee:jone:caff:bass:2014}.  

\subsection{Stopping Criterion}

Determining how long to run an MCMC simulation is critical to performing legitimate posterior inference. Premature termination often runs the risk of getting inaccurate estimates. The relative standard deviation fixed-width stopping rule (FWSR) \citep[see e.g.][]{fleg:gong:2015, gong:fleg:2015} is implemented to terminate the MCMC simulation. It is a member of the FWSR family \citep[for e.g. see][]{fleg:hara:jone:2008, jone:hara:caff:neat:2006, fleg:gong:2015}. The relative standard deviation FWSR is theoretically valid in that it terminates a simulation w.p.\ 1 and the resulting confidence interval achieves the nominal coverage probability. Moreover, it automates the stopping procedure for practitioners and outperforms convergence diagnostics in various numerical studies. Interested readers are directed to their papers for more details.

In short, the relative standard deviation FWSR terminates the simulation when the computational uncertainty is relatively small to the posterior uncertainty. Specifically, it controls the width of a confidence interval from a Markov chain central limit theorem through a threshold $\epsilon$ and significant level $\delta$. \cite{gong:fleg:2015} also establish a connection between the standard deviation FWSR and using effective sample size (ESS) as a stopping criteria, i.e. $K = 4 z_{\delta/2}^2 / \epsilon^2$, where $K$ is the number of effective samples and $z_{\delta/2}$ is a critical value from the standard Normal distribution. Based on this connection, for instance, setting $\epsilon = 0.124$ and $\delta = 0.05$ in the relative standard deviation FWSR is equivalent to terminate the simulation when an ESS reaches $K = 1000$. 

\subsection{Simulation Study} \label{sec:simu}

We report the results of a simulation study undertaken to validate the model and estimation procedure. The simulated dataset consists of a control group and five treatment groups. The control group is simulated based on estimates from maximum likelihood estimation (MLE) of a LMM on the control group of the experimental data. That is, denote $Y_{i,t,99}$ as the weight of mouse $i \in \{1, \cdots, 297\}$ from the control group (Diet No.99) taken at time $t \in \{365, 395, 456, 517, 578, 639, 700, 760, 821, 882, 943, 1004, 1065,1125, 1186\}$, we consider the following LMM based on \eqref{eq:lme},
\begin{equation} \label{eq:simuc}
Y_{i, t, 99} = \alpha_0 + \alpha_1 t + b_{0, i} + b_{1, i} t + \epsilon_{i, t}, \hspace{4mm} \epsilon_{i, t} \sim N (0, \sigma^2),
\end{equation}
where $\alpha_0$ and $\alpha_1$ are the global intercept and slope, $b_{0, i}$ and $b_{1, i}$ are the subject specific random effects, where $\pmb{b}_i = (b_{0, i}, b_{1, i} )^{T} \sim N_2 ( \pmb{0}, \lambda_D^{-1} I)$, and $\epsilon_{i, t}$ is the measurement error. Note that, as mentioned, the $\beta$'s in \eqref{eq:lme} are set to zero for the control group to serve as the baseline model.

Parameter estimation of \eqref{eq:simuc} was carried out using the lmer() function in the R package \texttt{lme4} \citep[][]{bate:maec:bolk:2012}. Notice time was rescaled using $t = (t - 365)/365$ prior to model fitting. The MLE estimates are $\pmb{\alpha} = (45.49, -5.75)^T$ and $\sigma^2 = 5.06$, and we set $\lambda_D^{-1} = 1.0$. Based on these estimates, we simulated 297 subjects from \eqref{eq:simuc} as the control group.

\begin{figure}[ht]
\centering
\includegraphics[width=0.6\textwidth]{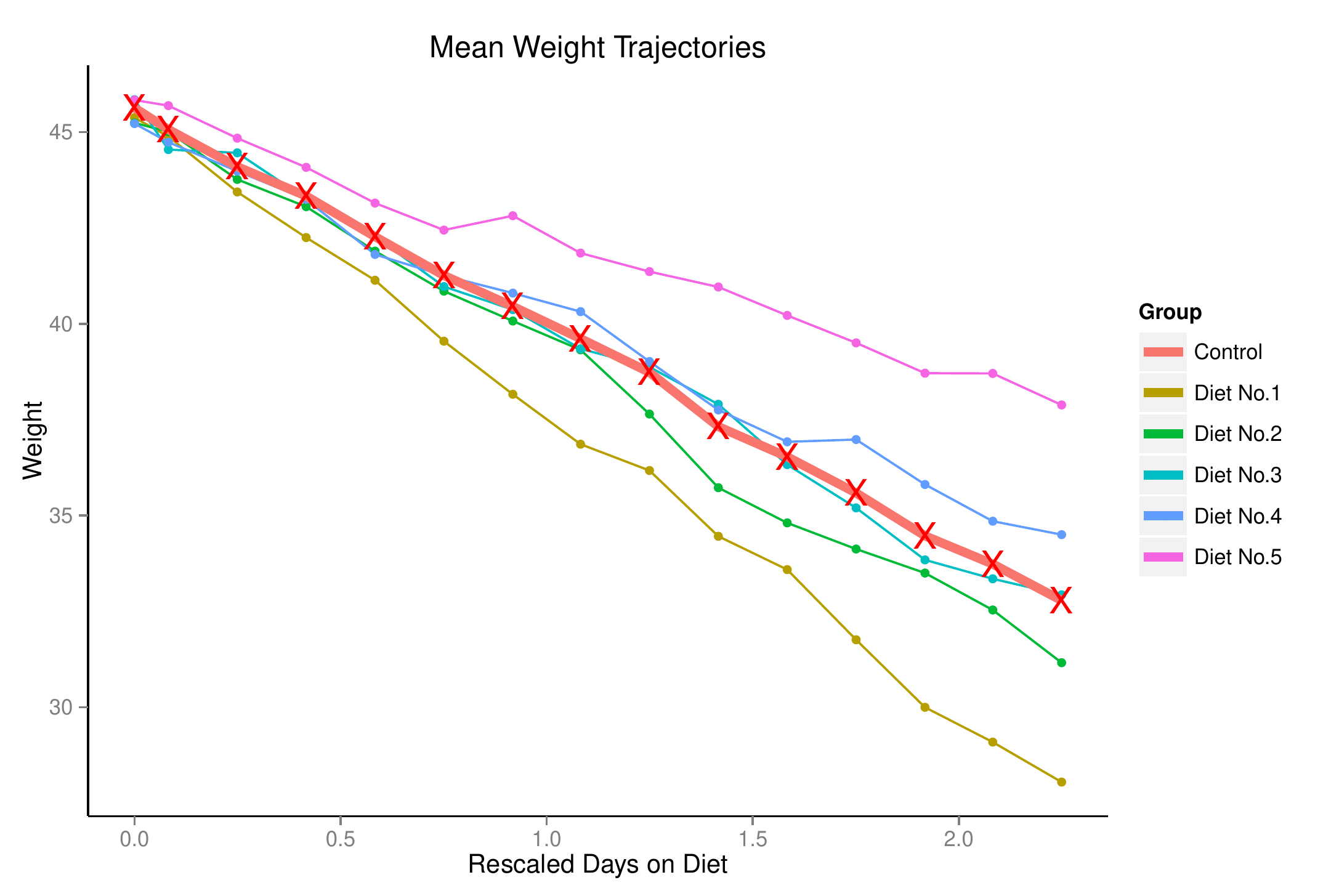}
\caption{Mean weight trajectories for the 5 treatment groups and the control group.}
\label{fig:simu}
\end{figure}

We then simulated five treatment groups, each with 36 subjects, by adding $\beta_{g, 1}$'s to \eqref{eq:simuc}, while keeping other settings the same as for the simulated control group,
\begin{equation} \label{eq:simut}
Y_{i, t, g} = \alpha_0 + \alpha_1 t + \beta_{g, 1} t + b_{0, i} + b_{1, i} t + \epsilon_{i, t}, \hspace{4mm} g = 1, \cdots, 5,
\end{equation}
where $\beta_{g, 1} \in \{-2.0, -0.5, 0.0, 0.5, 2.0 \}$ for each group. To be consistent with the experimental settings, we artificially differentiate the slope of each treatment group by $\alpha_1 + \beta_{g, 1}$, but maintained the same global intercept $\alpha_0$, since all mice were on the same diet at $t = 0$. Note that, we did not incorporate the ``die-off" mechanism from the experiment into the simulation. Figure~\ref{fig:simu} shows the mean weight trajectories for this simulated dataset. 

We followed the prior specification outlined in Section~\ref{sec:prior}. The hyper-parameters $d_1, d_2$ were set to $d_1 = 0.001, d_2=0.001$ for the prior on $\lambda_D$ to be vague. The hyper-parameters for $\pmb{\alpha} | \pmb{d}_3, d_4$ were set using estimates obtained from a fitted LMM on the simulated control group. The prior inclusion probabilities for the treatment groups, i.e. $\pi_g$, $g = 1, \cdots, 5$, were set to 0.5 for equal probability between inclusion and exclusion.%, which reflects no prior information on the treatment groups. 

The component-wise Gibbs sampler was run as described in Section~\ref{sec:posterior}. The simulation was terminated by the relative standard deviation FWSR with the tuning parameters $\epsilon = 0.124$ and $\delta = 0.05$. It resulted in 16385 iterations with an effective sample size of at least 1000 for estimation of the posterior mean of all the parameters. The resulting MCMC outputs show the chain is mixing well and centered near the true parameter values.

%\begin{figure}[ht]
%\centering
%%\includegraphics[width=0.8\textwidth]{var}
%\includegraphics[width=0.8\textwidth]{var-print}
%\caption{Gibbs sampler for variance terms $\lambda_D^{-1}$ and $\sigma^2$ in the simulation study.}
%\label{fig:var}
%\end{figure}

\begin{table}[ht]
\caption{Fixed-effects estimates for the simulated dataset. For MLE, mean and 95\% confidence interval (CI) are presented. For posterior inference, posterior mean, 95\% credible interval (CI) and marginal inclusion probability (standard error in the parenthesis) are presented.}
\begin{center}
\begin{tabular}{ccccccc}
\hline
\hline
 & & \multicolumn{2}{c}{MLE} & \multicolumn{3}{c}{Posterior} \\\cmidrule(lr){3-4} \cmidrule(lr){5-7}
Parameter & Truth & Mean & 95\% CI & Mean & 95\% CI & Pr($\gamma_{g, j} = 1 | y$) \\  
\hline
$\alpha_{0}$ & 45.50 & 45.570 & (45.431, 45.570) & 45.591 & (45.484, 45.700) &\\
$\alpha_{1}$ & -5.75 & -5.708 & (-5.852, -5.565) & -5.716 & (-5.822, -5.612) &\\
$\beta_{1, 1}$ & -2.00 & -2.130 & (-2.546, -1.713)  & -2.126 & (-2.562, -1.685) & 0.992(6.10e-5)\\
$\beta_{2, 1}$ & -0.50 & -0.693 & (-1.109, -0.276) & -0.698 & (-1.126, -0.267) & 0.983(7.44e-4)\\
$\beta_{3, 1}$ & 0.00 & -0.092 & (-0.508, 0.325) & -0.093 & (-0.518, 0.341) & 0.442(3.88e-3)\\
$\beta_{4, 1}$ & 0.50 & 0.708 & (0.292, 1.125) & 0.708 & (0.283, 1.136) & 0.987(5.74e-4)\\
$\beta_{5, 1}$ & 2.00 & 2.266 & (1.849, 2.683) & 2.268  & (1.830, 2.695) & 0.992(6.10e-5)\\
\hline
\end{tabular}
\label{tab:means}
\end{center}
\end{table} 

We compare our results to the estimates from the frequentist approach \citep[see e.g.][]{fitz:2004}, as it is widely used to model such problems. Researchers often combine LMM with certain model selection criteria, e.g. BIC, to determine which treatment are significantly differ from the control \citep[see e.g.][]{spin:mote:2013b}. Table~\ref{tab:means} contains the posterior estimates from the proposed model, along with the MLE estimates from \texttt{lme4} for the LMM. Despite that both approaches result in estimates close to the truth, our approach introduces the marginal inclusion probability for each treatment group that is vital to straightforward interpretation and correct ranking of the significance of the treatment effects. If a dichotomous decision is desired, setting a threshold to the suggested value of 0.8772, one would correctly classify Diet No.3 to be insignificant relative to the control.

The sensitivity to the prior inclusion probabilities was also evaluated by repeating the simulation with $\pi_g$'s set to ranging from 0.3 to 0.7. We found no difference in model ranking, although the parameter estimates were slightly different. Other simulation settings showed comparable parameter estimations between our method and the MLEs, and correctness in model ranking, although the results are not shown here.

\section{Application} \label{sec:app}

\begin{figure}
\centering
	\begin{subfigure}[t]{0.45\textwidth}
	\centering
	\includegraphics[width=\textwidth]{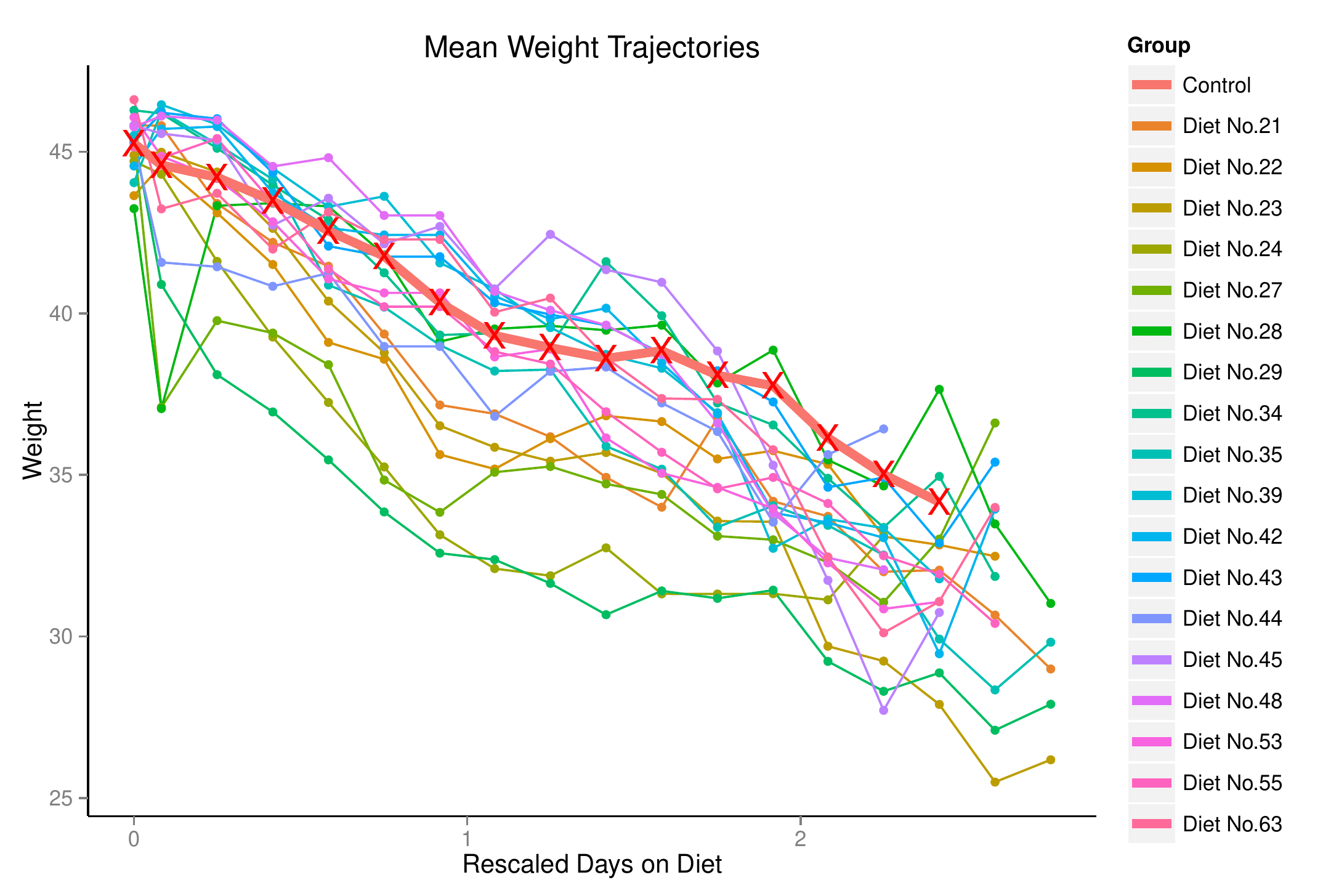}
	\caption{}
	\label{fig:app}
	\end{subfigure}%
	\hfill
	\begin{subfigure}[t]{0.45\textwidth}
	\centering
	\includegraphics[width=\textwidth]{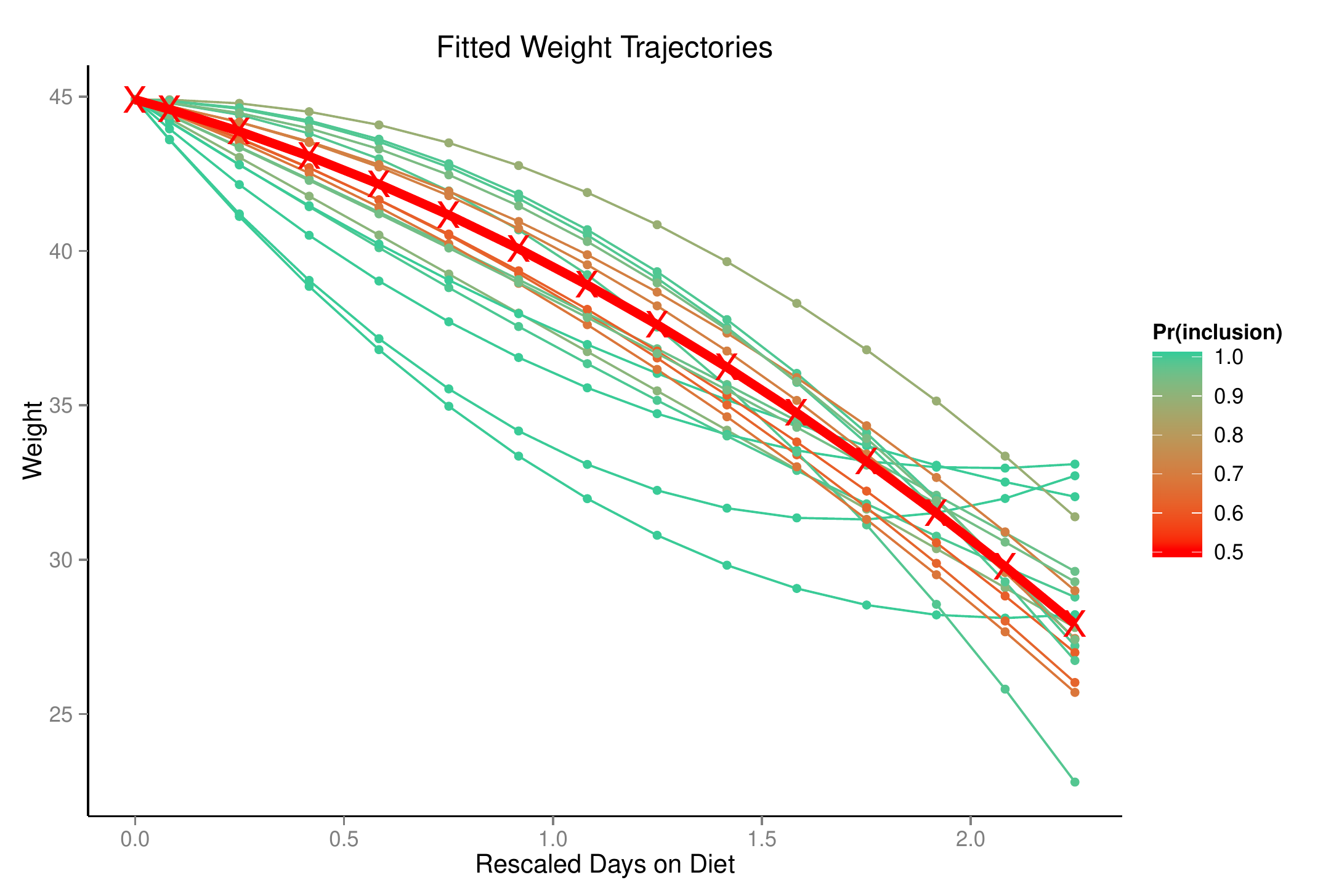}
	\caption{}
	\label{fig:comb-color}
	\end{subfigure}
\caption{Weight trajectories for the 18 treatment diets and the control diet.}
\label{fig:traj}
\end{figure}

In this section, we use the methodology detailed in Section~\ref{sec:model} to analyze the mouse body weight data (see Section~1.1). Out of the 56 treatment groups in the original study, we limited our attention to 18 pre-screened treatments that the researchers are most interested in, as well as the control diet (Diet No.99). The 18 groups exhibited altered weight trajectories, or were related chemically to groups that did.  Since all the groups consumed the same number of calories, weight trajectories are related to the disposition of dietary calories between body mass and metabolic energy, a key determinant of lifespan.
%Therefore, one would expect most of these diets are significantly different from the control diet. 
For simplicity, we denote these 19 diets as $\mathcal{G} = \{21, 22, 23, 24, 27, 28, 29, 34, 35, 39, 42, 43, 44, 45, 48, 53, 55, 63, 99 \}$. Similar to Section~\ref{sec:simu}, the days on diet were rescaled prior to analysis. Figure~\ref{fig:app} shows the mean weight trajectories for the 19 diet groups. Note that the mean weight estimates become unreliable as days on diet increases since mice died off in the process.

%\begin{figure}[ht]
%\centering
%\includegraphics[width=0.45\textwidth]{app.pdf}
%\caption{Mean weight trajectories for the 18 treatment diets and the control diet.}
%\label{fig:app}
%\end{figure}

As previously pointed out, since the subjects were on the same diet at $t = 0$, it is reasonable to assume the same intercept for all groups. The individual weight trajectories suggest that, unlike the simulated dataset, a quadratic term is needed to characterize the trajectories. Specifically, we re-write the LMM from \eqref{eq:ssvs} as
\begin{equation} \label{eq:app}
Y_{i, t, g} = \alpha_0 + \alpha_1 t + \alpha_2 t^2 + \beta_{g, 1} (\gamma_{g, 1}) t + \beta_{g, 2} (\gamma_{g, 2}) t^2 + b_{0, i} + b_{1, i} t + b_{2, i} t^2 + \epsilon_{i, t}, \hspace{4mm} g \in \mathcal{G}.
\end{equation}
Given the quantity of the data, it is possible to propose more complex models with additional polynomial terms. However, the additional terms complicate the model and the additional coefficients are difficult to interpret scientifically. Hence, we considered the model at \eqref{eq:app} as our full model.

Priors were specified as in the simulation study and $\pmb{\gamma}_{99}$ was set to $\pmb{\gamma}_{99} = (0, \cdots, 0)^T$ as the baseline model. The Gibbs sampler was terminated by the relative standard deviation FWSR with $\epsilon = 0.124$ and $\delta = 0.05$, resulting in 115792 iterations with at least 1000 effective samples for estimation of posterior mean of the parameters related to fixed-effects and variance components. The resulting MCMC outputs show the chain is mixing well and indicate that the variation among subjects outweighs the variance of the measurement errors. 

Table~\ref{tab:means-mouse} contains the posterior estimates from the proposed model, along with the MLE estimates from the R package \texttt{lme4} for the LMM. The results from the proposed method are comparable to the MLEs in terms of point and interval estimates. However, the marginal inclusion probability from the proposed method provides a direct measure of the significance for each diet, which was unavailable in previous investigations using the frequentist approach. 

%\begin{figure}[ht]
%\centering
%%\includegraphics[width=0.8\textwidth]{var-mouse}
%\includegraphics[width=0.8\textwidth]{var-mouse-print}
%\caption{Gibbs sampler for variance terms $\lambda_D^{-1}$ and $\sigma^2$ in the experimental application.}
%\label{fig:var-mouse}
%\end{figure}

\begin{table}[ht]
\caption{Fixed-effects estimates for the experimental dataset. For MLE, mean and 95\% confidence interval (CI) are presented. For posterior inference, posterior mean, 95\% credible interval (CI) and marginal inclusion probability (standard error in the parenthesis) are presented.}
\label{tab:means-mouse}
\fontsize{12}{12}\selectfont
\centering
\begin{tabular}{cccccc}
\hline
\hline
 & \multicolumn{2}{c}{MLE} & \multicolumn{3}{c}{Posterior} \\\cmidrule(lr){2-3} \cmidrule(lr){4-6}
Parameter & Mean & 95\% CI & Mean & 95\% CI & Pr($\gamma_{g, j} = 1 | y$)  \\ 
\hline
$\alpha_{0}$ & 44.879 & (44.638, 45.120) & 44.903 & (44.720, 45.086) &\\
$\alpha_{1}$ &  -4.124 & (-4.864, -3.384) & -3.687 & (-4.134, -3.237) &\\ 
$\alpha_{2}$ &  -0.992 & (-1.444, -0.540) & -1.718 & (-2.076, -1.369) &\\ 
\hline
$\beta_{21, 1}$ & -3.671 & (-6.637, -0.705)  & -3.800 & (-6.202, -1.432) & 0.991(2.13e-4)\\
$\beta_{21, 2}$ & 1.244 & (-0.553, 3.041)  & 1.667 & (-0.151, 3.475) & 0.809(1.15e-3)\\ 
\hline
$\beta_{22, 1}$ & -7.894 & (-10.828, -4.959) & -8.069 & (-10.353, -5.789) & 0.996(1.22e-5)\\
$\beta_{22, 2}$ & 4.429 & (2.644, 6.214) & 4.611 & (2.780, 6.461) & 0.996(8.63e-6)\\ 
\hline
$\beta_{23, 1}$ & -4.467 & (-7.282, -1.652) & -4.902 & (-6.995, -2.810) & 0.996(1.73e-5) \\
$\beta_{23, 2}$ & 1.819 & (0.203, 3.435) & 2.352 & (0.638, 4.046) & 0.970(4.75e-4) \\ 
\hline
$\beta_{24, 1}$ & -12.373 & (-15.317, -9.428) & -12.453 & (-14.795, -10.103) & 0.996(8.64e-6) \\
$\beta_{24, 2}$ & 5.349 & (3.533, 7.165) & 5.596 & (3.752, 7.444) & 0.996(8.64-e6) \\ 
\hline
$\beta_{27, 1}$ & -6.643 & (-8.694, -4.592) & -5.135  & (-6.671, -3.570) & 0.996(8.64e-6) \\
$\beta_{27, 2}$ & 3.173 & (1.948, 4.397) & 3.098  & (1.872, 4.344) & 0.996(8.64e-6) \\
\hline
$\beta_{28, 1}$ & 2.717 & (0.585, 4.849)  & 3.899 & (2.227, 5.611) & 0.997(1.50e-5) \\
$\beta_{28, 2}$ & -0.969 & (-2.273, 0.335)  & -1.047 & (-2.423, 0.316) & 0.747(1.27e-3) \\ 
\hline
$\beta_{29, 1}$ & -13.462 & (-15.574, -11.350) & -12.333 & (-13.946, -10.704) & 0.996(8.64e-6) \\
$\beta_{29, 2}$ & 6.499 & (5.195, 7.803) & 6.432 & (5.097, 7.751) & 0.996(8.64e-6) \\ 
\hline	
$\beta_{34, 1}$ & -0.768 & (-2.898, 1.363) & -0.892 & (-2.441, 0.638) & 0.663(1.39e-3) \\
$\beta_{34, 2}$ & 0.351 & (-1.002, 1.704) & 0.022 & (-1.315, 1.459) & 0.592(1.44e-3) \\ 
\hline
$\beta_{35, 1}$ & -1.552 & (-3.607, 0.503) & -2.355 & (-3.911, -0.752) & 0.983(3.33e-4) \\
$\beta_{35, 2}$ & 0.563 & (-0.670, 1.796) & 1.384 & (0.140, 2.657) & 0.920(7.83e-4) \\ 
\hline
$\beta_{39, 1}$ & 3.183 & (0.955, 5.410) & 2.699 & (0.829, 4.532) & 0.975(4.29e-4) \\
$\beta_{39, 2}$ & -2.422 & (-3.877, -0.967) & -2.212 & (-3.707, -0.659) & 0.976(4.16e-4) \\
\hline
$\beta_{42, 1}$ & 3.625 & (1.560, 5.689)  & 3.478 & (1.874, 5.054) & 0.996(2.86e-5) \\
$\beta_{42, 2}$ & -2.047 & (-3.298, -0.795)  & -1.686 & (-2.953, -0.423) & 0.963(5.28e-4) \\ 
\hline
$\beta_{43, 1}$ & 1.627 & (-0.442, 3.697) & 1.298 & (-0.201, 2.902) & 0.815(1.13e-3) \\
$\beta_{43, 2}$ & -0.659 & (-1.910, 0.591) & -0.364 & (-1.705, 0.929) & 0.584(1.45e-3) \\ 
\hline
$\beta_{44, 1}$ & -2.230 & (-4.532, 0.068) & -1.028 & (-2.777, 0.626) & 0.675(1.37e-3) \\
$\beta_{44, 2}$ & 0.853 & (-0.626, 2.331) & 0.274 & (-1.195, 1.889) & 0.549(1.46e-3) \\ 
\hline
$\beta_{45, 1}$ & 1.094 & (-1.122, 3.310) & 1.357 & (-0.351, 3.196) & 0.772(1.23e-3) \\
$\beta_{45, 2}$ & -0.160 & (-1.631, 1.310) & -0.696 & (-2.308, 0.875) & 0.633(1.41e-3) \\ 
\hline
$\beta_{48, 1}$ & 4.118 & (2.032, 6.204) & 3.361  & (1.695, 5.014) & 0.996(6.29e-5) \\
$\beta_{48, 2}$ & -2.268 & (-3.541, -0.995) & -1.728 & (-3.020, -0.439) & 0.960(5.48e-4) \\
\hline
$\beta_{53, 1}$ & -1.138 & (-3.242, 0.966) & -1.381 & (-2.955, 0.185) & 0.813(1.14e-3) \\
$\beta_{53, 2}$ & 0.047 & (-1.343, 1.249) & 0.176 & (-1.181, 1.574) & 0.530(1.47e-3) \\ 
\hline
$\beta_{55, 1}$ & -1.714 & (-3.800, 0.372) & -2.442  & (-4.072, -0.812) & 0.983(3.27e-4) \\
$\beta_{55, 2}$ & 0.695 & (-0.584, 1.975) & 1.356 & (0.061, 2.658) & 0.894(8.92e-4) \\
\hline
$\beta_{63, 1}$ & -2.483 & (-4.590, -0.376) & 2.712 & (1.052, 4.376) & 0.991(2.07e-4) \\
$\beta_{63, 2}$ & -1.367 & (-2.678, -0.056) & -1.305  & (-2.580, 0.019) & 0.863(1.00e-3) \\
\hline
\end{tabular}
\end{table}

\begin{figure}
\centering
	\begin{subfigure}[t]{0.45\textwidth}
	\centering
	\includegraphics[width=\textwidth]{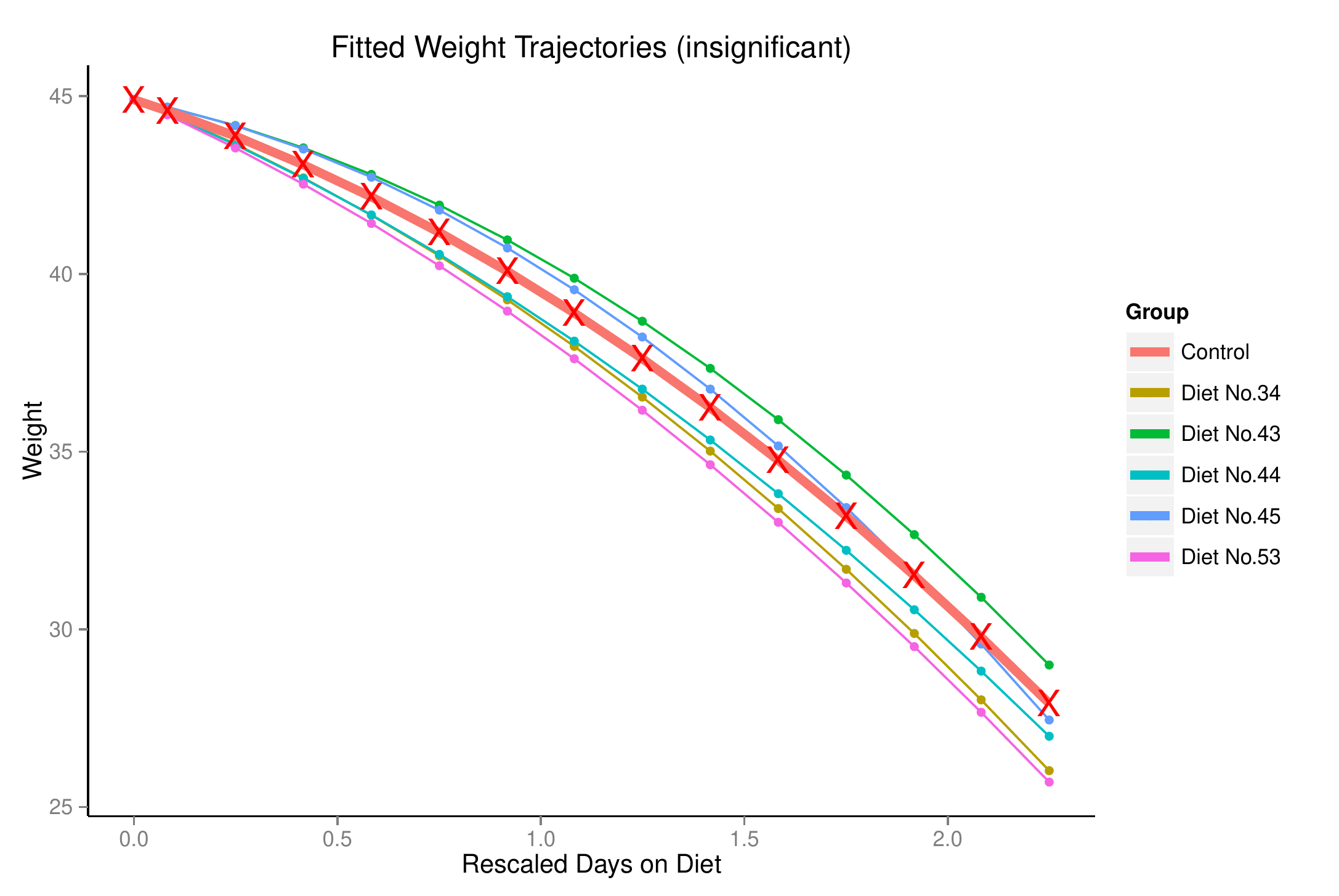}
	\caption{Estimated weight trajectories for treatment diets classified into insignificant and the control diet.}
	\label{fig:ana-insig}
	\end{subfigure}%
	\hfill
	\begin{subfigure}[t]{0.45\textwidth}
	\centering
	\includegraphics[width=\textwidth]{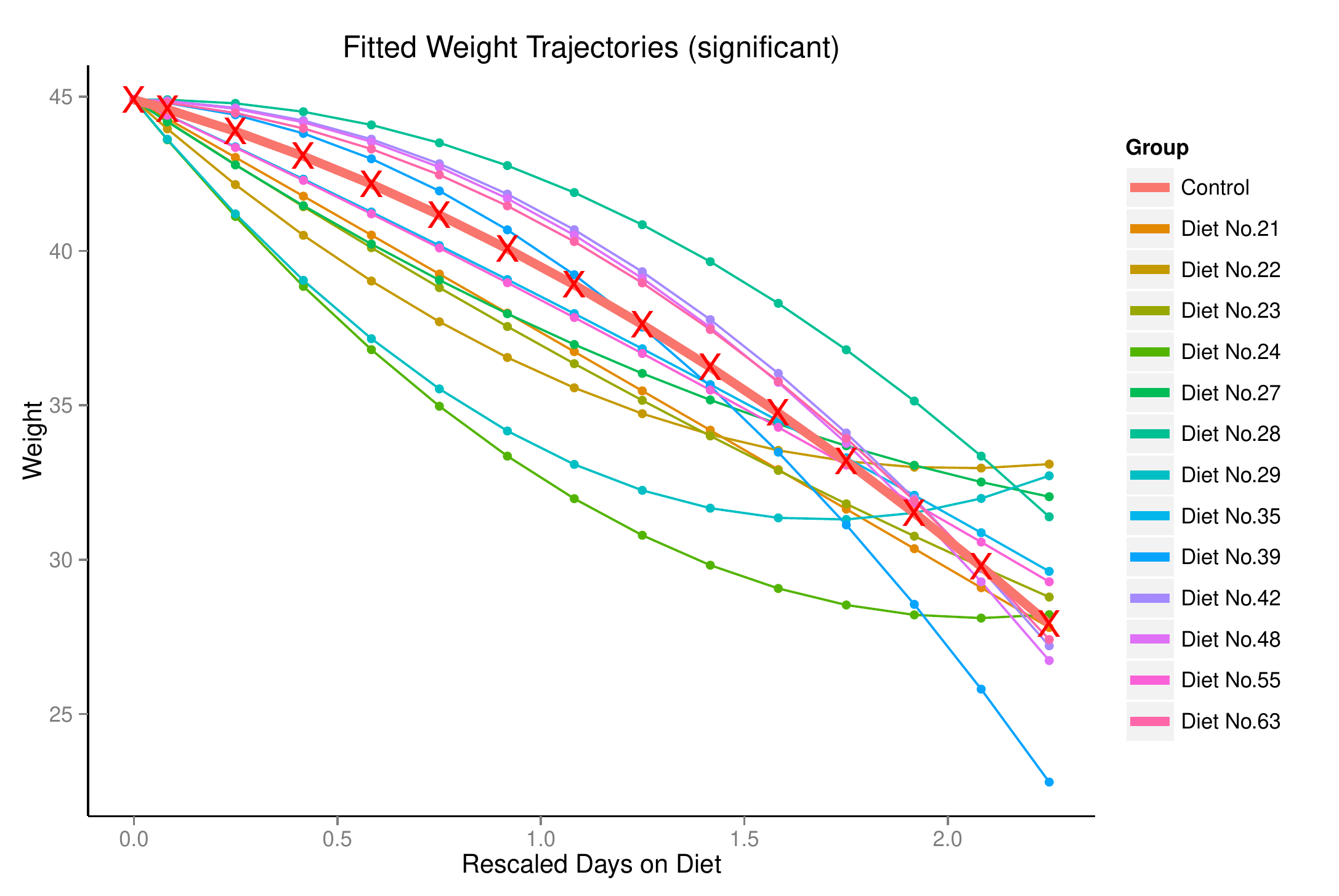}
	\caption{Estimated weight trajectories for treatment diets classified into significant and the control diet.}
	\label{fig:ana-sig}
	\end{subfigure}
	
	\bigskip 
	
	\begin{subfigure}[t]{0.45\textwidth}
	\centering
	\includegraphics[width=\textwidth]{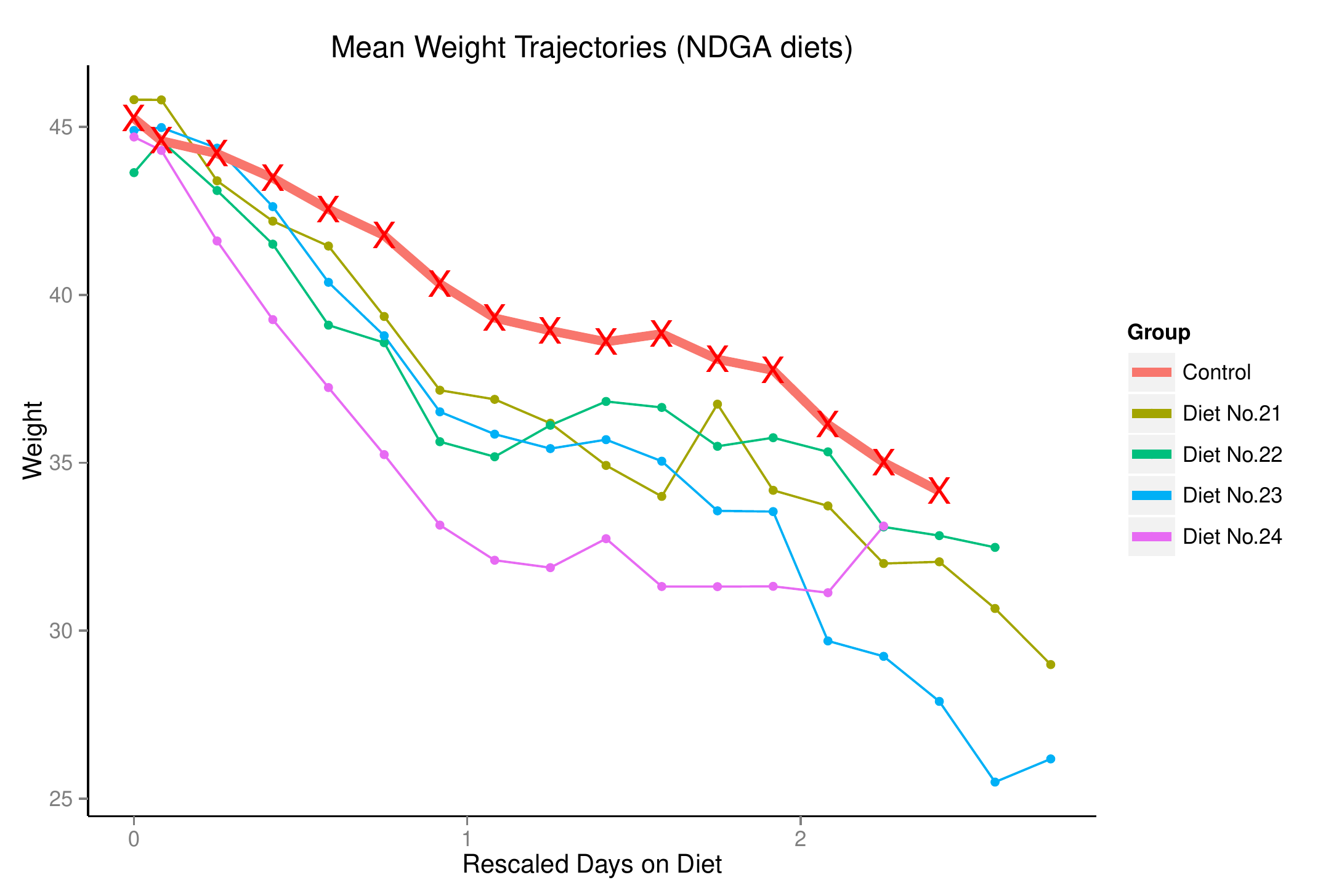}
	\caption{Mean weight trajectories for 4 NDGA supplemented diets and the control diet.}
	\label{fig:ana-mw}
	\end{subfigure}%
	\hfill
	\begin{subfigure}[t]{0.45\textwidth}
	\centering
	\includegraphics[width=\textwidth]{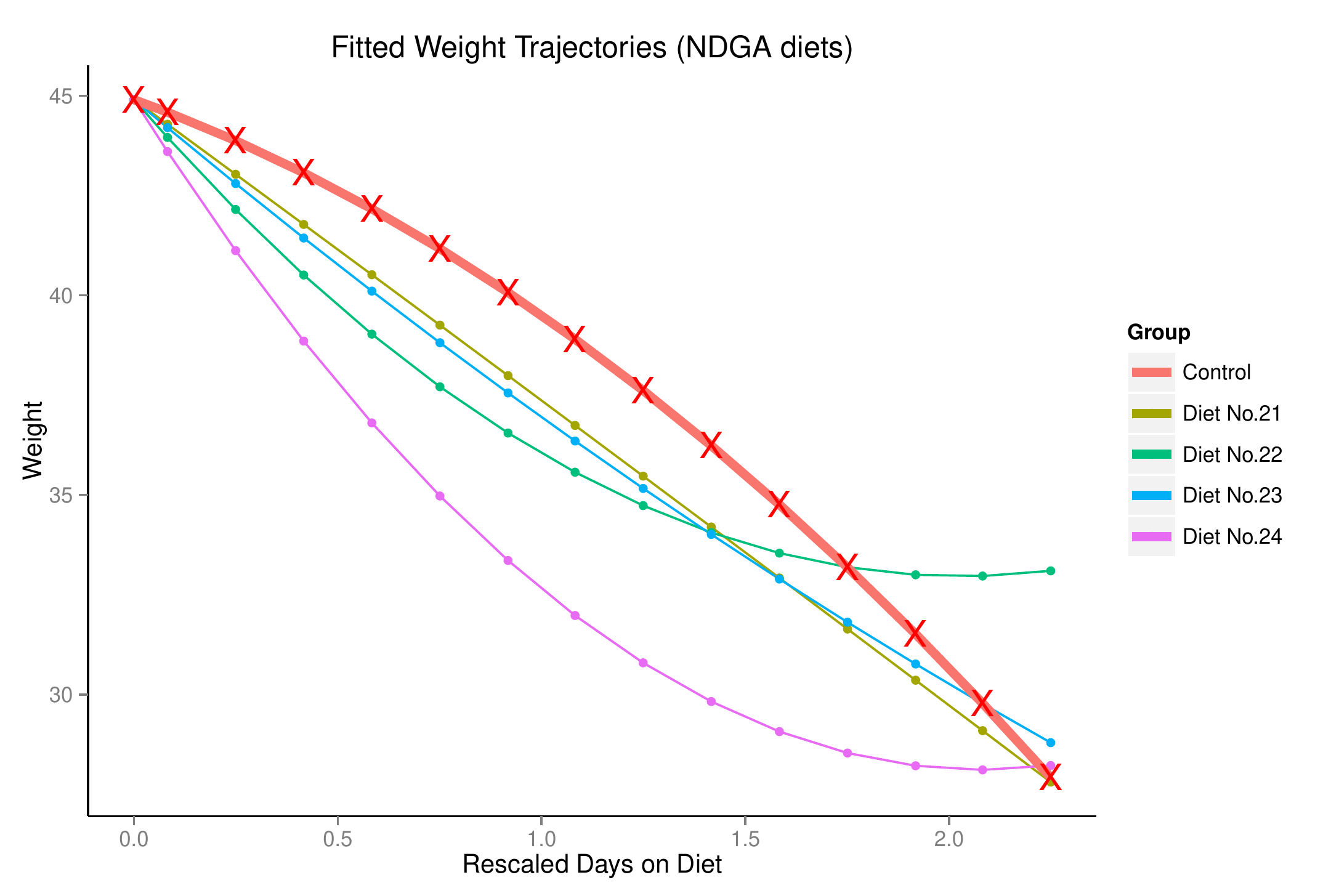}
	\caption{Estimated weight trajectories for 4 NDGA supplemented diets and the control diet.}
	\label{fig:ana-NDGA}
	\end{subfigure}
\caption{Analysis of the experimental dataset based on the proposed model.}
\label{fig:ana}
\end{figure}

Figure~\ref{fig:comb-color} shows the fitted weight trajectories colored based on the magnitude of their marginal inclusion probabilities. We can see that the treatment groups that have weight trajectories similar to the control group are the ones with lower inclusion probabilities. It shows that the marginal inclusion probability behaves well as a measure of the difference between a treatment diet and the control diet. Moreover, we find the suggested threshold 0.8772 is a reasonable value to classify treatment diets into significantly/insignificantly different from the control diet (see Figure~\ref{fig:ana-insig} and Figure~\ref{fig:ana-sig}). 

%\begin{figure}[ht]
%\centering
%\includegraphics[width=0.45\textwidth]{combined-color.pdf}
%\caption{Estimated weight trajectories for the 18 treatment diets and the control diet.}
%\label{fig:comb-color}
%\end{figure}

%\begin{figure}[ht]
%\centering
%\includegraphics[width=0.8\textwidth]{combined-insig-color.pdf} 
%\includegraphics[width=0.8\textwidth]{combined-sig-color.pdf}
%\caption{Estimated weight trajectories classified into significant/insignificant based on a threshold value of 0.8772.}
%\label{fig:comb-sig}
%\end{figure}

As an example, we compare four diets (Diet No.21 - Diet No.24) that are supplemented with 1.5, 2.5, 3.5, or 4.5 Nordihydroguaiaretic Acid (NDGA)/kg diet \citep[][]{spin:mote:2014c}. Figure~\ref{fig:ana-mw} shows the mean weight trajectories and Figure~\ref{fig:ana-NDGA} shows the fitted weight trajectories for the 4 diets supplemented with NDGA and the control diet. We find that the fitted trajectories correctly capture the characteristics of the mean weight trajectories for each diet, especially for the first half of the experiment when most mice were alive. These actual and fitted weight trajectories indicate that NDGA produced a dose-responsive decrease in body weight in the absence of a change in food consumption. These data suggest NDGA may have extend mouse lifespan by decreasing calorie absorption, inducing a state of caloric restriction, or by increasing metabolic rate. 
%The drugs present in diets 53, 27 and 29 also reduced body weight in the absence of a change in caloric consumption, suggesting they also affected the absorption or utilization of dietary calories.  
Further experiments will be required to resolve these possibilities.
%Also, Figure~\ref{fig:comb-sig} suggests that these four diets are significantly different from the control diet, and the mean weight trajectories confirm the level of significance indicated by the marginal inclusion probabilities in Table~\ref{tab:means-mouse}, e.g. Diet No.24 appears to be the most significantly different from the control diet.

%\begin{figure}[ht]
%\centering
%\includegraphics[width=0.8\textwidth]{mw-NDGA.pdf}
%\includegraphics[width=0.8\textwidth]{NDGA.pdf}
%\caption{Mean weight trajectories and the estimated weight trajectories for 4 NDGA related diets and the control diet.}
%\label{fig:NDGA}
%\end{figure}

\section{Discussion} \label{sec:disc}

This article proposes a novel method for Bayesian variable selection on LMM to compare multiple treatments with a control. It is built upon a modification of the fractional prior proposed by \cite{smit:kohn:1997} and a component-wise Gibbs sampler. It provides practitioners with a framework to incorporate prior knowledge of each treatment, as well as an intuitive evaluation of its significance. This method is quite general and has a wide range of potential applications in fields such as biology and medicine.

The proposed method is advantageous in that multiple treatments are compared to a control group simultaneously. In addition, the Bayesian framework introduces marginal inclusion probabilities for each group that allow direct measure of the significance of each treatment, which is difficult using alternative frequentist approaches. Notice that, we introduce a vector $\pmb{\gamma}_g = (\gamma_{g, 0}, \cdots, \gamma_{g, p-1})^T$ in \eqref{eq:ssvs} as indicator instead of a single $\gamma_g$ for each group, because it allows a more in-depth comparison between two groups. In this paper, the application on the experimental dataset provides new insights for researchers to group and study the diets based on their levels of significance. 

We emphasize careful posterior inference when using MCMC methodology. One major challenge for practitioners is determining how long to run a simulation. While some simulations are so complex that a fixed time approach is the only practical one, this is not so for most experiments. We advocate the use of relative standard deviation FWSR \citep[][]{fleg:gong:2015, gong:fleg:2015}, since it is proved to be easy to use, theoretically valid and superior to using convergence diagnostics as a stopping criteria \citep[][]{fleg:hara:jone:2008, jone:hara:caff:neat:2006}.

\section*{Acknowledgements}
The second author's work is partially supported by NSF grant DMS-13-08270.

\begin{appendix}
\section{Appendix} \label{app:post}

Full conditional posterior distributions are derived from \eqref{eq:post}, for $i = 1, \cdots, n$, $g = 1, \cdots, G$, and $j = 0, \cdots, p-1$. To calculate the full conditional posterior $q ( \gamma_{g, j} | \pmb{\alpha}, \pmb{\gamma}_{- (g, j)}, b, \sigma^2, y )$, we integrate out $\beta$ in \eqref{eq:post} as \cite{smit:kohn:1996},
\begin{equation} \label{eq:nobeta}
\begin{split}
&q(\pmb{\alpha}, \gamma, b, \sigma^2, \lambda_D | y) = \int q(\pmb{\alpha}, \beta, \gamma, b, \sigma^2, \lambda_D | y) d\beta \\
& \propto \left[ \prod_{g=1}^G \bigg[ \prod_{i \in g} \pi(\pmb{b}_i | \lambda_D) \bigg] \pi(\pmb{\gamma}_g) \right] \pi(\pmb{\alpha}) \pi(\lambda_D) \pi(\sigma^2) \\
& \qquad \times \prod_{g=1}^G \int_{\pmb{\beta}_g} \prod_{i \in g} p(\pmb{y}_i | \pmb{\alpha}, \pmb{\beta}_g, \pmb{\gamma}_g, \pmb{b}_i, \lambda_D, \sigma^2) \pi(\pmb{\beta}_g | \pmb{\alpha}, \pmb{\gamma}_g, b, \sigma^2) d \pmb{\beta}_g \\
\end{split}
\end{equation}
To calculate \eqref{eq:nobeta}, define $\pmb{\phi}_i = \pmb{y}_i - W_i \pmb{\alpha} - Z_i \pmb{b}_i$. For a given $g$, consider
\begin{equation*}
\begin{split}
&\int_{\pmb{\beta}_g} \prod_{i \in g} p(\pmb{y}_i | \pmb{\alpha}, \pmb{\beta}_g, \pmb{\gamma}_g, \pmb{b}_i, \lambda_D, \sigma^2) \pi(\pmb{\beta}_g | \pmb{\alpha}, \pmb{\gamma}_g, b, \sigma^2) d \pmb{\beta}_g \\
& \propto \int_{\pmb{\beta}_g} \left[ \prod_{i \in g} \sigma^{-n_i} \exp \left\{ - {1 \over 2\sigma^2} \bigg(\pmb{\phi}_i - X_i(\pmb{\gamma}_g) \pmb{\beta}_g(\pmb{\gamma}_g) \bigg)^T \bigg(\pmb{\phi}_i - X_i(\pmb{\gamma}_g) \pmb{\beta}_g(\pmb{\gamma}_g)\bigg) \right\}  \right] \\
& \quad \times |{1 \over \sigma^2} \sum_{i \in g} {1\over n_i} X_i^T(\pmb{\gamma}_g) X_i (\pmb{\gamma}_g) |^{1\over2} \times (2\pi)^{-{1\over2} \sum_{j=0}^{p-1} \gamma_{g, j}} \\
& \quad \times \exp \Bigg\{ -{1\over2} \left[ \pmb{\beta}_g(\pmb{\gamma}_g) - \bigg(\sum_{i \in g} {1\over n_i} X_i^T(\pmb{\gamma}_g) X_i(\pmb{\gamma}_g)\bigg)^{-1} \bigg(\sum_{i \in g} {1\over n_i} X_i^T(\pmb{\gamma}_g)\pmb{\phi}_i \bigg) \right]^T \bigg({1\over \sigma^2} \sum_{i \in g} X_i^T(\pmb{\gamma}_g) X_i(\pmb{\gamma}_g) \bigg) \\
& \hspace{6em} \left[ \pmb{\beta}_g(\pmb{\gamma}_g) - \bigg(\sum_{i \in g} {1\over n_i} X_i^T(\pmb{\gamma}_g) X_i(\pmb{\gamma}_g)\bigg)^{-1} \bigg(\sum_{i \in g} {1\over n_i} X_i^T(\pmb{\gamma}_g)\pmb{\phi}_i \bigg) \right] \Bigg\} d\pmb{\beta}_g \\
& = \sigma^{-\sum_{i \in g} n_i} {\left( | \sum_{i \in g} {1\over n_i} X_i^T(\pmb{\gamma}_g) X_i(\pmb{\gamma}_g) | \over | \sum_{i \in g} (1 + {1\over n_i}) X_i^T(\pmb{\gamma}_g) X_i(\pmb{\gamma}_g) | \right) }^{1\over2} \\
& \quad \times  \exp \Bigg\{ -{1\over 2\sigma^2} \sum_{g=1}^G  \Bigg[ \sum_{i \in g} \pmb{\phi}_i^T \pmb{\phi}_i + \bigg(\sum_{i\in g} {1\over n_i} X_i^T(\pmb{\gamma}_g) \pmb{\phi}_i\bigg)^T \bigg(\sum_{i\in g} {1 \over n_i} X_i^T(\pmb{\gamma}_g) X_i(\pmb{\gamma}_g)\bigg)^{-1} \bigg(\sum_{i\in g} {1\over n_i} X_i^T(\pmb{\gamma}_g) \pmb{\phi}_i\bigg) \\
& \hspace{4em} -\bigg(\sum_{i\in g} (1 + {1 \over n_i})  X_i^T(\pmb{\gamma}_g) \pmb{\phi}_i\bigg)^T \bigg(\sum_{i\in g} (1 + {1 \over n_i}) X_i^T(\pmb{\gamma}_g) X_i(\pmb{\gamma}_g)\bigg)^{-1} \bigg(\sum_{i\in g} (1 + {1 \over n_i})  X_i^T(\pmb{\gamma}_g) \pmb{\phi}_i\bigg) \Bigg]  \Bigg\}
\end{split}
\end{equation*}
Therefore, \eqref{eq:nobeta} is further simplified
\begin{equation} \label{eq:nobetafinal}
\begin{split}
& = \lambda_D^{np/2} \exp \{ -{\lambda_D \over 2} \sum_{g=1}^G\sum_{i \in g} \pmb{b}_i^T \pmb{b}_i \} \prod_{g=1}^G \prod_{j=1}^p \pi_g^{\gamma_{g, j}}(1-\pi_g)^{1 - \gamma_{g, j}} \exp \{ -{1\over2} (\pmb{\alpha} - \pmb{d}_3)^T d_4 (\pmb{\alpha} - \pmb{d}_3) \} \\
& \hspace{5mm} \times \lambda_D^{d_1 - 1} \exp \{ -d_2\lambda_D \} \prod_{g=1}^G \sigma^{-\sum_{i \in g} n_i} {\left( | \sum_{i \in g} {1\over n_i} X_i^T(\pmb{\gamma}_g) X_i(\pmb{\gamma}_g) | \over | \sum_{i \in g} (1 + {1\over n_i}) X_i^T(\pmb{\gamma}_g) X_i(\pmb{\gamma}_g) | \right) }^{1\over2} \\
& \quad \times  \exp \Bigg\{ -{1\over 2\sigma^2} \sum_{g=1}^G  \Bigg[ \sum_{i \in g} \pmb{\phi}_i^T \pmb{\phi}_i + \bigg(\sum_{i\in g} {1\over n_i} X_i^T(\pmb{\gamma}_g) \pmb{\phi}_i\bigg)^T \bigg(\sum_{i\in g} {1 \over n_i} X_i^T(\pmb{\gamma}_g) X_i(\pmb{\gamma}_g)\bigg)^{-1} \bigg(\sum_{i\in g} {1\over n_i} X_i^T(\pmb{\gamma}_g) \pmb{\phi}_i\bigg) \\
& \hspace{4em} - \bigg(\sum_{i\in g} (1 + {1 \over n_i})  X_i^T(\pmb{\gamma}_g) \pmb{\phi}_i\bigg)^T \bigg(\sum_{i\in g} (1 + {1 \over n_i}) X_i^T(\pmb{\gamma}_g) X_i(\pmb{\gamma}_g)\bigg)^{-1} \bigg(\sum_{i\in g} (1 + {1 \over n_i})  X_i^T(\pmb{\gamma}_g) \pmb{\phi}_i\bigg) \Bigg]  \Bigg\}
\end{split}
\end{equation}
Based on \eqref{eq:nobetafinal}, the full posterior distribution is characterized by
\begin{equation*}%\label{eq:gamma}
\begin{split}
& q ( \gamma_{g, j} | \pmb{\alpha}, \pmb{\gamma}_{- (g, j)}, b, \sigma^2, y ) \propto  \pi_{g}^{\gamma_{g, j}} ( 1- \pi_g)^{1- \gamma_{g, j}} \left( {|\sum_{i \in g} {1\over n_i} X_i^T (\pmb{\gamma}_g) X_i (\pmb{\gamma}_g) | \over |\sum_{i \in g} (1 + {1\over n_i}) X_i^T (\pmb{\gamma}_g) X_i (\pmb{\gamma}_g)} \right)^{1\over2} \\
& \times \exp \Bigg\{ -{1\over 2\sigma^2} \Bigg[ \sum_{i \in g} \pmb{\phi}_i^T \pmb{\phi}_i + \bigg(\sum_{i\in g} {1\over n_i} X_i^T(\pmb{\gamma}_g) \pmb{\phi}_i\bigg)^T \bigg(\sum_{i\in g} {1 \over n_i} X_i^T(\pmb{\gamma}_g) X_i(\pmb{\gamma}_g)\bigg)^{-1} \bigg(\sum_{i\in g} {1\over n_i} X_i^T(\pmb{\gamma}_g) \pmb{\phi}_i\bigg) \\
& \hspace{4em} - \bigg(\sum_{i\in g} (1 + {1 \over n_i})  X_i^T(\pmb{\gamma}_g) \pmb{\phi}_i\bigg)^T \bigg(\sum_{i\in g} (1 + {1 \over n_i}) X_i^T(\pmb{\gamma}_g) X_i(\pmb{\gamma}_g)\bigg)^{-1} \bigg(\sum_{i\in g} (1 + {1 \over n_i})  X_i^T(\pmb{\gamma}_g) \pmb{\phi}_i\bigg) \Bigg]  \Bigg\},
\end{split}
\end{equation*}
where $\pmb{\gamma}_{- (g, j)} = (\gamma_{g, 0}, \cdots, \gamma_{g, j-1}, \gamma_{g, j+1}, \cdots, \gamma_{g, p-1})^T$.

\end{appendix}

\setstretch{1}
\bibliographystyle{apalike}
\bibliography{ref}

\end{document}